\documentclass[prb,aps,superscriptaddress,twocolumn]{revtex4}

\usepackage{graphicx}
\usepackage{dcolumn}
\usepackage{bm}
\usepackage{amssymb}

\begin{document}


\title{Quantum criticality of vanadium chains with strong relativistic spin-orbit interaction}

\author{Gia-Wei~Chern}
\affiliation{Department of Physics, University of Wisconsin,
Madison, Wisconsin 53706, USA}
\author{Natalia~Perkins}
\affiliation{Department of Physics, University of Wisconsin,
Madison, Wisconsin 53706, USA}
\author{George~I.~Japaridze}
\affiliation{Andronikashvili Institute of Physics, Tamarashvili str.
6, 0177 Tbilisi, Georgia} \affiliation{Ilia State University,
Colokashvili Avenue 3-5, 0162 Tbilisi, Georgia }


\begin{abstract}
We study quantum phase transitions induced by the on-site spin-orbit
interaction $\lambda \mathbf L\cdot\mathbf S$ in a toy model of
vanadium chains. In the $\lambda\to 0$ limit, the decoupled spin and
orbital sectors are described by a Haldane and an Ising chain,
respectively. The gapped ground state is composed of a ferro-orbital
order and a spin liquid with finite correlation lengths. In the
opposite limit, strong spin-orbital entanglement results in a
simultaneous spin and orbital-moment ordering, which can be viewed
as an orbital liquid. Using a combination of analytical arguments
and density-matrix renormalization group calculation, we show that
an intermediate phase, where the ferro-orbital state is accompanied
by a spin N\'eel order, is bounded on both sides by Ising transition
lines. Implications for vanadium compounds CaV$_2$O$_4$ and
ZnV$_2$O$_4$ are also discussed.
\end{abstract}

\maketitle

Quasi-one-dimensional Mott insulators with strongly coupled spin and
orbital degrees of freedom have attracted considerable attention
recently. A well-studied case is the Kugel-Khomskii Hamiltonian with
an SU(2) symmetry in both  spin and orbital sectors. \cite{li98}
This model is believed to describe the essential physics of quasi-1D
compounds Na$_2$Ti$_2$Sb$_2$O and NaV$_2$O$_5$. \cite{axtell,isobe}
Extensive numerical and analytical studies have revealed a rich
phase diagram. \cite{itoi,azaria} Of particular interest is a SU(4)
symmetric point of the Hamiltonian where the low-energy physics is
described by a conformal field theory with a central charge $c=3$,
equivalent to a model of three free bosons.

\begin{figure} [t]
\includegraphics[width=0.99\columnwidth]{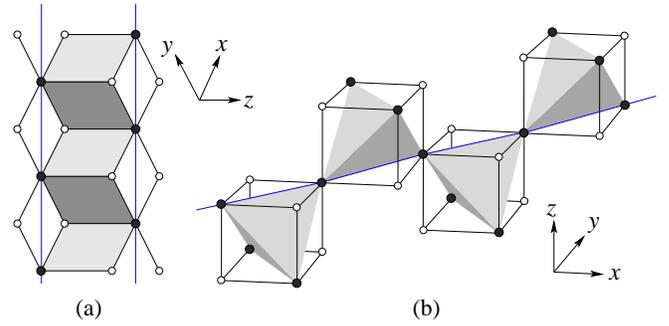}
\caption{\label{fig:v-chains} Vanadium chains in (a) CaV$_2$O$_4$
and (b) ZnV$_2$O$_4$. The black and white circles denote vanadium
and oxygen ions, respectively. The V$^{3+}$ ions are arranged in a
zigzag chain of edge-sharing VO$_6$ octahedra in CaV$_2$O$_4$. On
the other hand, V$^{3+}$ ions in spinel form a pyrochlore lattice,
which can be viewed as a cross-linking network of vanadium chains.
The quasi-1D spin-1 chains are highlighted by solid blue lines.}
\end{figure}

In this paper, we investigate an 1D spin-orbital system which in
many aspects is different from the above SU(2)$\times\,$SU(2) model.
The interest is partly motivated by recent experimental progresses
on several vanadates including spinel ZnV$_2$O$_4$
\cite{reehuis,lee} and quasi-1D CaV$_2$O$_4$. \cite{pieper,niazi}
The vanadium chains in these compounds are characterized by
frustrated magnetic interactions, Ising-like orbital exchanges, and
a large relativistic spin-orbit (SO) interaction. The origin of the
first two features can be traced to the lattice geometry of these
compounds: vanadium ions in spinel form a three-dimensional
pyrochlore lattice, and in CaV$_2$O$_4$ they are arranged in weakly
coupled zigzag chains (Fig.~\ref{fig:v-chains}). In both structures
the 90$^\circ$ angle between vanadium-oxygen bonds in a network of
edge-sharing VO$_6$ octahedra makes direct exchange the primary
mechanism for inter-site spin-orbital interaction. An important
consequence is that the spin-orbital Hamiltonian only depends on
orbitals through the corresponding projection operators. For
example, direct $dd\sigma$ exchange takes place along a $[110]$ bond
only when one or both of the $d_{xy}$ orbitals are occupied, giving
rise to an Ising-like orbital interaction. \cite{dimatteo,chern09a}

\begin{figure}
\centering
\includegraphics[width=0.9\columnwidth]{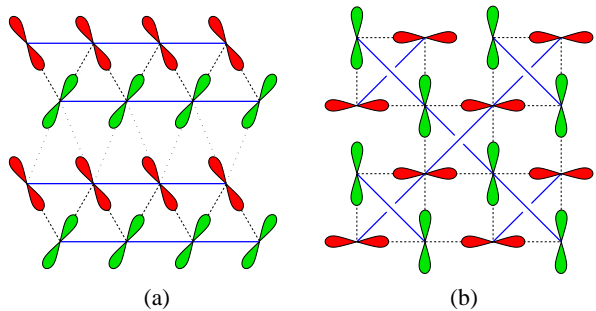}
\caption{\label{fig:orbitals} Schematic diagram of antiferro-orbital
orders in (a) CaV$_2$O$_4$ and (b) ZnV$_2$O$_4$. In both cases, the
$d_{xy}$ orbital is occupied at all sites. The interaction between
the remaining $d_{zx}$ and $d_{yz}$ orbitals (indicated by red and
green symbols, respectively) is governed by an antiferromagnetic
Ising-like interaction on the dashed bonds. Also note the {\em
ferro-orbital} order along the quasi-1D spin-1 chains (solid blue
lines).}
\end{figure}

Contrary to the anisotropic orbital exchange, magnetic interaction
governed by the Heisenberg Hamiltonian preserves the spin-rotational
symmetry. The spin exchange constant, however, depends on the
underlying orbital occupations. Combined with geometrical
frustration, the orbital-dependent spin exchange renders these
vanadates essentially quasi-1D spin systems. To see this, we note
that the local $d_{xy}$ orbitals are always occupied due to
distorted VO$_6$ octahedron in both compounds. As a result, the
largest spin-spin interaction takes place on bonds parallel to
$[110]$ and $[1\bar 10]$ directions in spinels
[Fig.~\ref{fig:v-chains}(b)], whereas for CaV$_2$O$_4$ the dominant
spin exchange occurs along the two rails of a zigzag chain
[Fig.~\ref{fig:v-chains}(a)]. Furthermore, couplings between these
spin chains are not only weak, but also geometrically frustrated.
\cite{tsunetsugu,ot} Since V$^{3+}$ ions in both compounds have spin
$S=1$, their magnetic properties thus can be understood from the
viewpoint of weakly coupled spin-1 chains (solid lines in
Fig.~\ref{fig:orbitals}).

In both vanadates, one of the two electrons of the V$^{3+}$ ion
always occupies the low-energy $d_{xy}$ state, the other one
occupies either $d_{yz}$ or $d_{zx}$ orbitals. We introduce a
pseudospin-$1/2$ operator $\bm\tau$ to describe this doublet orbital
degeneracy, where $\tau^a$ are the Pauli matrices. We choose a basis
such that $\tau^z = \pm 1$ corresponds to states $|yz\rangle$ and
$i|zx\rangle$, respectively. The dominant pseudospin interaction is
governed by an antiferromagnetic Ising-like Hamiltonian on
nearest-neighbor bonds connecting different spin-1 chains
\cite{dimatteo,chern09a} (dashed lines in Fig.~\ref{fig:orbitals});
the relevant orbitals on these bonds are $d_{yz}$ and $d_{zx}$. Due
to the static and three-dimensional nature of the orbital Ising
Hamiltonian, the system tends to first develop a long-range orbital
order upon lowering the temperature. Fig.~\ref{fig:orbitals} shows a
schematic diagram of the 3D antiferro-orbital order for the two
compounds. It is important to note that orbitals on individual
spin-1 chains (solid lines in Fig.~\ref{fig:orbitals}) are {\em
ferromagnetically} ordered.

The antiferro-orbital order shown in Fig.~\ref{fig:orbitals}(b),
however, is incompatible with the observed crystal symmetry
$I4_1/amd$ of tetragonal ZnV$_2$O$_4$. \cite{reehuis,lee} The
discrepancy can be attributed to a large SO interaction $\lambda
\mathbf L\cdot\mathbf S$ of vanadium ions. Indeed, the SO term is
minimized by a state with simultaneous N\'eel ordering of spins and
orbital angular momenta. \cite{ot,chern09b} The corresponding
orbital order consisting of complex $d_{yz}\pm i d_{zx}$ orbitals
preserves both the mirror inversion $m$ and diamond glide $d$, and
is consistent with the experimental data.

In the absence of SO interaction, the ground state of the spin-1
chain is a nondegenerate spin singlet. This spin liquid phase, also
known as the Haldane phase, must be separated from the N\'eel state
favored by a large SO coupling by quantum phase transitions. Since a
nonzero $L^x$ requires the electron be in a complex orbital state
$\frac{1}{\sqrt{2}}\left(|zx\rangle \pm i |xy\rangle\right)$, a
fully occupied $d_{xy}$ orbital thus results in the vanishing of
$L^x$ and $L^y$. \cite{dimatteo,ot} The nonzero $z$ component of the
orbital angular momentum is given by $L^z = - \tau^x$ in our
representation. To understand the critical behavior of vanadium
chains due to the $LS$ coupling, we consider the following
spin-orbital Hamiltonian:
\begin{eqnarray}
    \label{eq:H0}
    {H} = J \sum_{n} \mathbf S_n\!\cdot\! \mathbf S_{n+1} - K
    \sum_{n} \tau^z_n \tau^z_{n+1} - \lambda \sum_n  \tau^x_n S^z_n.
\end{eqnarray}
This simple model describes two well-studied 1D systems, i.e. an
$S=1$ Haldane chain and a ferromagnetic Ising chain (both $J$, $K >
0$), coupled together by an on-site SO interaction $\lambda\,
\mathbf L\cdot\mathbf S$. Note that the eigenstates of $\tau^x$,
$|yz\rangle \pm i |zx\rangle$, carry an angular momentum $L^z = \mp
1$, respectively. The Hamiltonian Eq.~(\ref{eq:H0}) has a
U(1)$\times$ $Z_2$$\times$$Z_2$ symmetry: the spin SU(2) symmetry is
reduced to $\mbox{U(1)}\times Z_2$ by the SO term, whereas an
additional $Z_2$ symmetry comes from the orbital Ising Hamiltonian.

If should be noted that the ferro-orbital order along the quasi-1D
chains is stabilized by the inter-chain antiferro-orbital coupling
in real compounds. The ferromagnetic exchange $-K$ in
Eq.~(\ref{eq:H0}) thus should be regarded as an effective coupling
in the mean-field sense.

Despite its simplicity, the model contains rather rich physics. It
is easy to see that the first-order correction vanishes identically
in the ground state of decoupled Haldane and Ising chains. To have a
glimpse of the effects of the SO interaction, one needs to go to
higher orders and examine the elementary excitations of
model~(\ref{eq:H0}). We start with the kink excitations of the Ising
chain. Kinks, or domain walls, are topological defects separating
the two degenerate ground states of perfectly aligned psedospins.
For classical Ising chains, kinks are static quasiparticles with a
constant energy $2 K$. To obtain the quasiparticle operators, we
first fermionize the Ising chain using Jordan-Wigner transformation
\cite{ising}
\begin{eqnarray}
    \label{JW}
    \tau^z_n = \prod_{m<n} \bigl(2c^\dagger_m c^{\phantom{\dagger}}_m-1
    \bigr) \bigl(c^{\phantom{\dagger}}_n + c^\dagger_n\bigr),
    \quad
    \tau^x_n = 1 - 2 c^\dagger_n c^{\phantom{\dagger}}_n.
\end{eqnarray}
The kink operator $\gamma_q$ is obtained after subsequent Fourier
and Bogoliubov transformations $\gamma_q = u_q c_q - i v_q
c^\dagger_{-q}$, where $u_q = \cos(q/2)$ and $v_q = \sin(q/2)$.
\cite{ising}

For spin-1 chain, the lowest excitation above the singlet ground
state is a triplet with a dispersion $\omega_k \approx [\Delta_0^2 +
v^2(k - \pi)^2]^{1/2}$ near the energy minimum. Here $v=2.56\, J$ is
the spinwave velocity and $\Delta_0 \approx 0.4 J$ is the Haldane
gap. \cite{huse} To model the low-energy physics of spin-1 chain, we
follow Ref.~\onlinecite{affleck90} and introduce three massive
magnons $a^{\pm}_k$ and $a^z_k$ carrying quantum number $S^z = \pm
1$, and 0, respectively. The spin operator is $S^z_n = l^z_n +
\frac{1}{2}\sum_k \phi_k\,e^{ikn}/\sqrt{L}$, where
\begin{eqnarray}
    \phi_k = \sqrt{\frac{2v}{\omega_{\pi+k}}} \bigl(a^z_{\pi+k} +
    a_{-\pi-k}^{z\dagger}\bigr),
\end{eqnarray}
and the uniform part $l^z_n$ is quadratic in transverse magnons
$a^{\pm}_k$. In terms of magnons and kinks,
Hamiltonian~(\ref{eq:H0}) becomes
\begin{eqnarray}
    & & {H} = \sum_{k,\sigma} \omega_k\, a^{\sigma\dagger}_k\, a^{\sigma\phantom{\dagger}}_k
    + 2 K \sum_q \gamma^\dagger_q \gamma^{\phantom{\dagger}}_q \\
    & & \quad +\frac{\lambda}{\sqrt{L}}\sum_{k, q, q'}\!\!'\bigl[u_{q+q'} \phi^{\phantom{\dagger}}_k
    \gamma^\dagger_q \gamma^{\phantom{\dagger}}_q
    + \frac{i}{2} v_{q-q'}
    \phi^{\phantom{\dagger}}_k\bigl(\gamma^\dagger_q
    \gamma^\dagger_{q'} - \gamma^{\phantom{\dagger}}_{q'} \gamma^{\phantom{\dagger}}_q\bigr)\bigr].
    \nonumber
\end{eqnarray}
The prime on the summation indicates conservation of momentum $k=q
\pm q'$. In obtaining the above expression, we have neglected the
interaction between kinks and transverse magnons $a^{\pm}_k$.

\begin{figure}
\centering
\includegraphics[width=0.75\columnwidth]{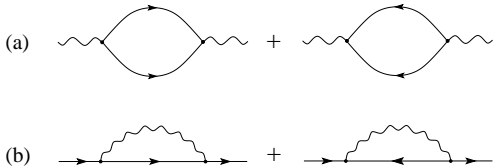}
\caption{\label{fig:1loop} One-loop corrections to self-energy of
(a) magnon and (b) kink. The solid and wavy lines denote the kink
and magnon propagators, respectively.}
\end{figure}

Assuming $\lambda \ll J, \,K$, we employ a perturbation theory to
examine the properties of quasiparticles in the presence of SO
coupling. We first evaluate the two-kinks bubbles shown in
Fig.~\ref{fig:1loop}(a):
\begin{eqnarray}
    \label{eq:pi1}
    \Pi_1(\omega) = \frac{2\lambda^2 K}
    {\omega^2 - 16 K^2},
\end{eqnarray}
It is interesting to note that the particle-hole bubble does not
contribute to $\Pi_1$. The above expression for $\Pi_1(\omega)$
diverges as the magnon energy approaches that of a pair of kinks,
i.e. $\omega \approx 4K$. In this regime magnons strongly interact
with the kinks, and higher-order corrections to the interaction
vertex have to be taken into account. Assuming $\omega \ll K$ and
using random-phase approximation to compute the magnon self-energy,
we obtain a renormalized spin gap
\begin{eqnarray}
    \label{eq:ds}
    \Delta_s \approx \Delta_0 - \frac{v\lambda^2}{4\Delta_0 K},
\end{eqnarray}
which decreases with increasing $\lambda$. At large $\lambda$,
closing of the spin gap indicates a phase transition into a spin
ordered phase characterized by $\langle a_{\pi} \rangle \neq 0$.

Fig.~\ref{fig:1loop}(b) shows the one-loop contribution to the
self-energy of magnons. In the $K \gg \Delta_0$ limit, the second
term in Fig.~\ref{fig:1loop}(b) is negligible compared with the
first one, we obtain a self-energy
\begin{eqnarray}
    \Sigma_1(q) \approx -
    \frac{\lambda^2}{\Delta_0} \bigl(1 + e^{-\Delta_0/v} \cos 2q\bigr)
\end{eqnarray}
The energy of the kink excitation given by $\varepsilon_q \approx 2K
+ \Sigma_1(q)$ indicates that the kinks become mobile through the
mediation of virtual magnons.

The perturbative calculation gives important insight to the
elementary excitations in the small $\lambda$ regime of the
spin-orbital model (\ref{eq:H0}). In particular, the reduced spin
and orbital gaps indicate quantum phase transitions at finite
$\lambda$. To investigate the nature of the phase transitions and
the properties of possible new phases, we numerically investigate
the spin-orbital model Eq.~(\ref{eq:H0}) using the infinite-system
density-matrix renormalization group (DMRG) method. \cite{white92}
The DMRG calculation is known to give a rather accurate description
of the ground-state properties for 1D systems. In our calculation we
have employed periodic boundary conditions in order to accommodate
the staggered ordering of spins and orbital angular momenta.

\begin{figure}
\begin{centering}
\includegraphics[height=0.58\columnwidth]{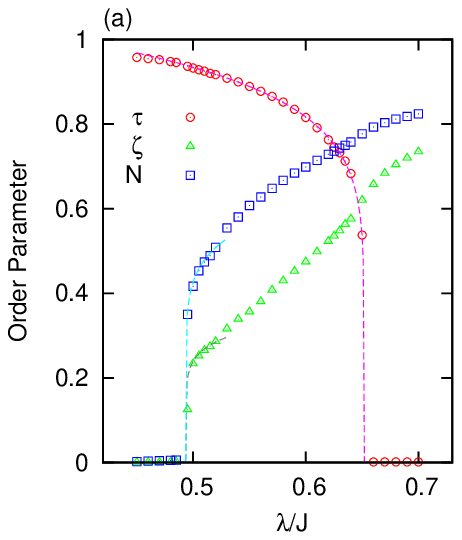}
\includegraphics[height=0.58\columnwidth]{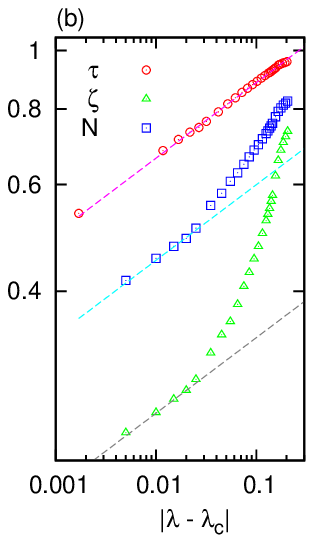}
\caption{\label{fig:dmrg} (a) Order parameters as a function of
$\lambda/J$ for a chain with $K = 0.5 J$. The dashed lines are
fittings to 2D Ising transition $m \sim |\lambda -
\lambda_c|^{1/8}$, where $m$ is the corresponding order parameter.
(b) The same plot in log-log scale.}
\end{centering}
\end{figure}

Noting that $L^z = -\tau^x$ in our representation, we define the
following order parameters:
\begin{eqnarray}
    \langle \tau^z_n \rangle = \tau, \quad
    \langle S^z_n \rangle = \mathcal{N}(-1)^n ,\quad
    \langle L^z_n \rangle =  -\zeta(-1)^n.
\end{eqnarray}
The numerical results of a spin-orbital chain with exchange
constants $K = 0.5 J$ are shown in Fig.~\ref{fig:dmrg}. At small
$\lambda$, the ground state is characterized by a nonzero
ferro-orbital order $\tau \neq 0$, while the spin sector is in the
disordered Haldane phase with $\mathcal{N} = 0$. At critical point
$\lambda_{c1} \simeq 0.491 J$, the linear chain undergoes a quantum
phase transition into a state with simultaneous ordering of
staggered spin and orbital-moment, characterized by nonzero order
parameters $\mathcal{N}$ and $\zeta$, respectively. The
ferro-orbital order $\tau$ remains finite in the intermediate phase.
As we further increase $\lambda$, the system undergoes yet another
quantum phase transition at $\lambda_{c2} \simeq 0.657 J$. This
critical point is marked by the melting of the ferro-orbital order
$\tau$. In the $\lambda \to \infty$ limit, both order parameters
$\mathcal{N},\,\zeta$ approach 1. The ground state of the
spin-orbital chain consists of alternatively occupied states
$|S^z=\pm 1\rangle\otimes|yz\rangle\mp i |zx\rangle$. The orbital
occupation numbers $n_{yz} = n_{zx} = 1/2$ are uniform along the
chain.

Since the order parameters $\mathcal{N}$ and $\tau$ describe
respectively the broken $Z_2$ symmetry of the spin and orbital
sectors, both critical points $\lambda_{c1}$ and $\lambda_{c2}$ are
expected to be in the 2D Ising universality class. Indeed, by
fitting the corresponding order parameter $m$ to the Ising scaling
relation $m\sim|\lambda - \lambda_c|^{1/8}$, we find agreeable
result as shown by the dashed lines in Fig.~\ref{fig:dmrg}. The
Ising nature of the spin N\'eel transition can be understood in the
limit of large orbital gap $K\gg J$. By integrating out orbitals,
one obtains an easy-axis spin anisotropy: $ -D S_z^{\,2}$, where $D
= \lambda^2/4K$; the resultant spin gap is reduced in accordance
with Eq.~(\ref{eq:ds}). As demonstrated numerically in
Ref.~\onlinecite{chen03}, the spin-1 chain undergoes an Ising
transition into a N\'eel state when $D > D_c$.

Above $\lambda_{c1}$, a nonzero $\mathcal{N}$ exerts an effective
(staggered) transverse field on the orbitals. By rotating
pseudospins an angle $\pi$ about $\tau^z$ axis on odd-numbered
sites, the orbital sector is described by a quantum Ising
Hamiltonian:
\begin{eqnarray}
    \label{eq:ising}
    H_{\rm orbital} = -K\sum_n \tau^z_n\tau^z_{n+1} - \Gamma\sum_n
    \tau^x_n,
\end{eqnarray}
where the transverse field $\Gamma = \lambda \mathcal{N}$. It is
known that the Ising chain reaches a critical state at $\Gamma_c=K$.
\cite{ising} Numerically, we obtain an N\'eel order $\mathcal{N}
\approx 0.659$ at critical point $\lambda_{c2}$. The corresponding
critical field $\Gamma_c =\lambda_{c2} \mathcal{N} \approx 0.52 J$
is indeed close to the orbital exchange $K = 0.5 J$ used in the DMRG
calculation.

\begin{figure}
\includegraphics[width=0.85\columnwidth]{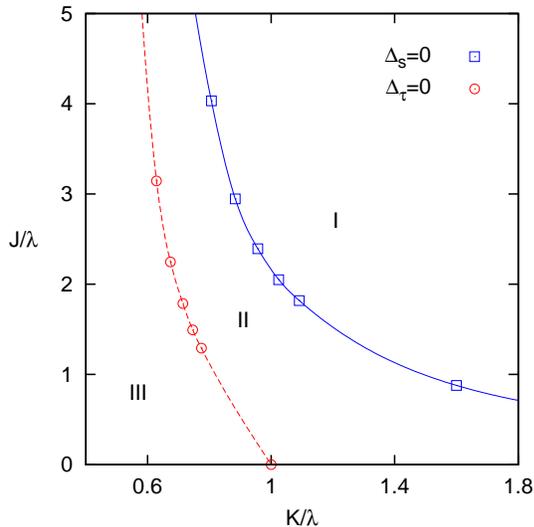}
\caption{\label{fig:phase} Phase diagram of spin-orbital
model~(\ref{eq:H0}). The lines are guide for the eye. In phase I,
the spin sector is in the disordered Haldane phase,
$\mathcal{N}=\zeta=0$, while the orbitals are ferromagnetically
ordered $\tau \neq 0$. The ground state of phase III is composed of
N\`eel spin order and orbital-moment order, which can be viewed as
an orbital liquid, i.e. $\mathcal{N}\neq 0$ and $\tau = 0$. In the
intermediate phase II, a spin N\'eel order coexists with the
ferro-orbital Ising order.}
\end{figure}

Our main results are summarized in the phase diagram
Fig.~\ref{fig:phase} which contains three massive phases separated
by two Ising transition lines. The spin sector in the $J = 0$ limit
is extensively degenerate as each spin could be in either $|S^z = +
1\rangle$ or $|S^z = -1\rangle$ states, independently; the total
degeneracy is $2^L$. After applying a $\pi$-rotation about $\tau^z$
axis to those pseudospins at sites where $S^z_n = -1$, the orbital
sector is again mapped to a quantum Ising chain Eq.~(\ref{eq:ising})
with the transverse field $\Gamma = \lambda$. Since the orbital
Ising transition occurs at $\Gamma_c = K$, the phase boundary
$\lambda_{c2}$ hence ends at $K = \lambda_{c2}$ on the $J = 0$ axis.
In the small $J$ limit, one can estimate $\lambda_{c1}$ for the spin
N\'eel transition using the critical condition $D = D_c \approx
0.05J$, \cite{chen03} which gives $JK \propto \lambda_{c1}^2$. The
region of intermediate phase enclosed by boundaries $\lambda_{c1}$
and $\lambda_{c2}$ shrinks with increasing spin exchange $J$. At
very large $J$, the two Ising lines could merge to form a Gaussian
criticality or a first-order transition.

We now discuss implications of our findings to vanadium compounds.
As discussed before, the ferro-Ising order parameter $\tau$ can also
serve as the 3D antiferro-orbital order parameter for both vanadates
(see Fig.~\ref{fig:orbitals}). A nonzero $\tau$ thus creates two
different orbital chains, hence further lowering the crystal
symmetry. However, experiments on both compounds observed a higher
symmetry \cite{reehuis,pieper} indicating that vanadium chains in
both vanadates are likely in the $\tau = 0$ orbital liquid state
(phase III). Furthermore, the appearance of finite orbital moment
$L^z = \zeta$ antiparallel to spin $S^z$ at $\lambda > \lambda_{c2}$
also explains the reduced vanadium moment $\mu = (2S^z + L^z)\mu_B
\approx 1\mu_B$ observed experimentally. \cite{lee,pieper} On a
final note, we caution that the fermionic description of orbital
excitations as kinks in an Ising chain is the consequence of using
1D approximation for the orbital system. Furthermore, the
Kugel-Khomskii-type spin-orbital terms $(\tau^z_i\tau^z_j)\,(\mathbf
S_i\cdot\mathbf S_j)$ introduce correlations between orbital and
spin excitations. A detailed description of these 3D spin-orbital
excitations will be discussed in future publications.

{\em Acknowledgments}. The authors are grateful to A.~Chubukov,
A.~Kolezhuk, O.~Tchernyshyov, and in particular A.A.~Nersesyan for
stimulating discussions. N.P. acknowledges the support from
NSF-DMR1005932 and the hospitality of visitors program at MPIPKS, where
part of the work on this manuscript has been done.
G. I. J. acknowledges the support from GNSF-ST09/4-447.

\end{document}